\documentclass[twocolumn]{aastex631}

\usepackage{graphicx}
\usepackage{amsmath}
\usepackage{gensymb}

\begin{document}

\title{No Evidence of Asymmetrically Enhanced Star Formation in Infalling Galaxies in UNIONS}

\correspondingauthor{Lauren M. Foster}
\email{fostel8@mcmaster.ca}

\author[0000-0003-0214-9194]{Lauren M. Foster}
\affiliation{Department of Physics and Astronomy, McMaster University, 1280 Main Street West, Hamilton, ON, L8S 3L8, Canada}

\author[0000-0003-4722-5744]{Laura C. Parker}
\affiliation{Department of Physics and Astronomy, McMaster University, 1280 Main Street West, Hamilton, ON, L8S 3L8, Canada}

\author[0000-0001-8221-8406]{Stephen Gwyn}
\affiliation{NRC Herzberg Astronomy \& Astrophysics, 5071 West Saanich Road, BC, V9E 2E7, Canada}

\author[0000-0002-0692-0911]{Ian D. Roberts}
\affiliation{Department of Physics \& Astronomy, University of Waterloo, 200 University Avenue West, Waterloo, ON, N2L 3G1, Canada}
\affiliation{Waterloo Centre for Astrophysics, University of Waterloo, 200 University Avenue West, Waterloo, ON, N2L 3G1, Canada}

\author[0000-0002-6639-4183]{James E. Taylor}
\affiliation{Department of Physics \& Astronomy, University of Waterloo, 200 University Avenue West, Waterloo, ON, N2L 3G1, Canada}
\affiliation{Waterloo Centre for Astrophysics, University of Waterloo, 200 University Avenue West, Waterloo, ON, N2L 3G1, Canada}

\author[0000-0002-1437-3786]{Michael J. Hudson}
\affiliation{Department of Physics \& Astronomy, University of Waterloo, 200 University Avenue West, Waterloo, ON, N2L 3G1, Canada}
\affiliation{Waterloo Centre for Astrophysics, University of Waterloo, 200 University Avenue West, Waterloo, ON, N2L 3G1, Canada}
\affiliation{Perimeter Institute of Theoretical Physics, 31 Caroline St. N., Waterloo, ON, N2L 2Y5, Canada}

\author[0000-0003-4666-6564]{Alan W. McConnachie}
\affiliation{NRC Herzberg Astronomy \& Astrophysics, 5071 West Saanich Road, BC, V9E 2E7, Canada}

\author[0000-0001-5486-2747]{Thomas de Boer}
\affiliation{Institute for Astronomy, University of Hawaii, 2680 Woodlawn Drive, Honolulu, HI, 96822, USA}

\begin{abstract}

Ram pressure stripping is a well-known environmental quenching mechanism that removes gas from galaxies infalling into groups and clusters. In some extreme examples of ram pressure stripping, galaxies with extended gas tails show evidence of enhanced star formation prior to quenching. In this work we use a sample of 5277 local satellite galaxies in which a stripped tail of gas has not necessarily been observed, to quantify the strength of ram pressure-enhanced star formation and compare these results to a control sample of 8360 field galaxies. We use $u$-band imaging from the Ultraviolet-Near Infrared Northern Survey (UNIONS) as a star formation tracer and several metrics to quantify star formation asymmetry. We compare these results to environmental properties of the galaxy, such as their time since infall and host halo mass, to constrain the degree of ram pressure enhanced star formation as a function of environment. We find no significant differences between the satellite and the field samples. We further restrict our sample to galaxies which we most expect to be experiencing significant ram pressure but find no strong evidence of these galaxies having systematically enhanced star formation. Finally, we investigate the properties of the most asymmetric galaxies in our sample and again find no strong evidence of ram pressure-induced star formation enhancement. We conclude that any star formation enhancement must be small for infalling galaxies, suggesting that this effect is either uncommon or short-lived.

\end{abstract}


\section{Introduction}

Galaxies are largely bimodal in many of their properties: they are either blue in colour, with high gas content, elevated star formation rates (SFRs), and late-type morphologies, or they show red colours, low gas content, low SFRs, and early-type morphologies \citep[e.g.][]{strateva_2001,brinchmann_2004,baldry_2006,blanton_2009,peng_2010}. Galaxies typically begin as star-forming, blue galaxies and transition to passive, red galaxies through quenching processes. The rarity of galaxies observed in this transitional stage suggests that quenching is a relatively rapid process \citep{salim_2014}.

There are numerous pathways to galaxy quenching, with many having a strong dependence on environment. Galaxies living in dense environments, such as groups and clusters, experience additional environmental effects that isolated field galaxies do not, making them less star-forming on average \citep[e.g.][]{wetzel_2012,haines_2015}. In dense environments, galaxies experience both gravitational and hydrodynamic interactions that isolated galaxies do not. Dynamical interactions include galaxy minor and major mergers, which can induce bursts of star formation that quickly deplete star-forming gas reservoirs leading to quenching \citep[e.g.][]{mihos_1994a,mihos_1994b}. In addition, tidal interactions are capable of stripping star-forming gas \citep[e.g.][]{mayer_2006,chung_2007}. 

In this work, we focus on ram pressure, a hydrodynamical environmental effect. Ram pressure stripping (RPS) is well-known as a quenching mechanism, as it can remove star-forming gas from galaxies falling into groups and clusters \citep[e.g.][]{gunn_1972,quilis_2000}. Ram pressure scales as $\rho_{\text{ICM}}v^2$, where $\rho_{\text{ICM}}$ is the density of the intracluster medium (ICM), and $v$ is the infall speed of the galaxy relative to the ICM \citep{gunn_1972}. Thus, ram pressure is strongest in galaxy clusters, where the ICM density is higher and galaxies typically have higher velocities. The circumgalactic medium and the outer atomic gas is most susceptible to stripping, as it is more loosely bound than the inner molecular gas, and is therefore removed first, while the existing stellar population is unaffected \citep{boselli_2022}.

Systems which are known to be experiencing ram pressure stripping and have observable tails of stripped gas opposing their direction of motion are often called jellyfish galaxies. Tails are most readily observable in H$\alpha$ \citep[e.g.][]{gavazzi_2001,yoshida_2002,sun_2007,yagi_2007,yagi_2010,poggianti_2017,boselli_2018}, HI \citep[e.g.][]{kenney_2004,oosterloo_2005,chung_2007,chung_2009,kenney_2015}, and the radio continuum \citep[e.g.][]{gavazzi_1995,roberts_2022}, but can be identified at many wavelengths such as infrared \citep[e.g.][]{sivanandam_2010,sivanandam_2014}, optical \citep[e.g.][]{gavazzi_1995,gavazzi_2001,sun_2007,sivanandam_2010,sun_2010,yagi_2010,fumagalli_2014,fossati_2016,gavazzi_2017,roberts_2020}, submillimeter \citep[e.g.][]{scott_2013,jachym_2014,jachym_2019}, UV \citep[e.g.][]{smith_2010}, and x-ray \citep[e.g.][]{sun_2006,sun_2010,sun_2022}. In many cases, it has been found that jellyfish galaxies have higher than normal star formation rates on their leading edge \citep[e.g.][]{gavazzi_2001,troncosoiribarren_2020,lee_2016,roberts_2022,hess_2022,zhu_2024}. It is hypothesized that the gas in the ISM on the leading edge of these galaxies is compressed by ram pressure, leading to the conversion of atomic to molecular gas, higher densities of molecular gas in the galaxy, and more efficient star formation \citep[e.g.][]{schulz_2001,moretti_2020a, moretti_2020b,cramer_2021}. A recent wind tunnel simulation of a low-mass infalling galaxy has also found that ram pressure can funnel gas towards the centres of galaxies, resulting in enhanced SFRs \citep{zhu_2024}. 

The evidence is strong that jellyfish galaxies have this enhanced leading-side star formation. However, since jellyfish are rare, it is unclear how universal enhanced star formation is for infalling satellite galaxies in groups and clusters. \cite{rodriguez_2020} tries to address this question by comparing the $g-r$ colours of the leading and trailing halves of infalling galaxies in SDSS, and find that under the right conditions, the leading half is bluer than the trailing half, consistent with ram pressure-induced star formation. \cite{troncosoiribarren_2020} use the EAGLE \citep{schaye_2015} simulations to compare the SFR on the leading and trailing halves, and find that ram pressure can enhance SFR by $\sim10\%$. However, they note that this enhancement would be difficult to detect observationally, where galaxies are viewed in projection and the true direction of motion is not known. It is currently unknown how common this phase of galaxy evolution is and if the star formation enhancement seen in many jellyfish galaxies is a special case.

In this work, we use $u$-band imaging from the Ultraviolet-Near Infrared Optical Northern Survey (UNIONS; Gwyn et al.~2024, in preparation; see also \citealt{ibata_2017}) as a star formation tracer and use several direction-dependent and independent measures sensitive to asymmetries in star formation. We compare a large sample of group and cluster galaxies to a control field sample and investigate the environmental dependence of star formation asymmetry. The outline of this paper is as follows: in Section \ref{section_2}, we describe the data used and the cuts made to create our samples. In Section \ref{section_3}, we describe all measurements made on the data, including each of our star formation asymmetry metrics. In Section \ref{section_4} we compare the star formation asymmetries of group and cluster galaxies to those of isolated galaxies and explore how these results depend on galaxy and environmental properties. In Section \ref{section_5} we discuss the significance of these results, and in Section \ref{section_6} we conclude with a summary of our findings. In this work, we assume a flat $\Lambda$CDM cosmology with $H_0=70\text{km s}^{-1}\text{Mpc}^{-1}$, $\Omega_{M}=0.3$, and $\Omega_\Lambda=0.7$.

\section{Data}
\label{section_2}

In this work, we use galaxies in the Sloan Digital Sky Survey Data Release 7 (SDSS DR7; \cite{abazajian_2009}) with spectroscopic redshifts matched to the deep $u$-band imaging of UNIONS. Here, we describe the relevant data and catalogs.

\subsection{Yang Group and Cluster Catalog}
To differentiate group, cluster, and field galaxies, we use the Yang Group Catalog for SDSS DR7 \citep{yang_2007}. This catalogue identifies galaxy groups using an iterative halo-based group finder algorithm and estimates the mass, extent, and velocity dispersion of the dark matter halo of each group. For this work, we use their dark matter halo mass derived from an $M_h-M_L$ relation and compute each group's virial radius and velocity dispersion using equations (5) and (6) in \cite{yang_2007}:

\begin{equation}
    r_{180}=1.26h^{-1}\text{Mpc}\left(\frac{M_h}{10^{14}h^{-1}M_\odot}\right)^{1/3}(1+z_{\text{group}})^{-1}
\end{equation}

\begin{equation}
    \sigma=397.9\text{km s}^{-1}\left(\frac{M_h}{10^{14}h^{-1}M_\odot}\right)^{0.3214}
\end{equation}

\noindent where $M_h$ is the dark matter halo mass, and $z_{\text{group}}$ is the redshift of the group centre. For further details, we refer the reader to \cite{yang_2007}.

We restrict our sample to satellite galaxies in host halos with $M_h>10^{13}M_\odot$ and number of group members $\geq3$. We define group galaxies to have $10^{13}M_\odot\leq M_h<10^{14}M_\odot$ and cluster galaxies to have $M_h\geq10^{14}M_\odot$. We expand upon the \cite{yang_2007} group and cluster membership by including all galaxies within the projected phase space (PPS) regions of interest (Section \ref{section_phasespace}). Galaxies in SDSS with $M_h<10^{13}M_\odot$ outside these PPS regions are considered field galaxies.

\subsection{GALEX-SDSS-WISE Legacy Catalog}
We use stellar mass and star formation rate (SFR) estimates taken from the GALEX-SDSS-WISE Legacy catalogue (GSWLC-M2; \cite{salim_2016,salim_2018}), derived using UV/optical/midIR SED fitting. We keep galaxies with stellar masses $M_*\geq10^{9}M_\odot$ (while the sample is complete to $M_*\sim10^{10.5}M_\odot$) as lower mass galaxies are the most susceptible to RPS \citep{gunn_1972,hester_2006,steyrleithner_2020,boselli_2022}. 

\begin{figure}
    \centering
    \includegraphics[width=\linewidth]{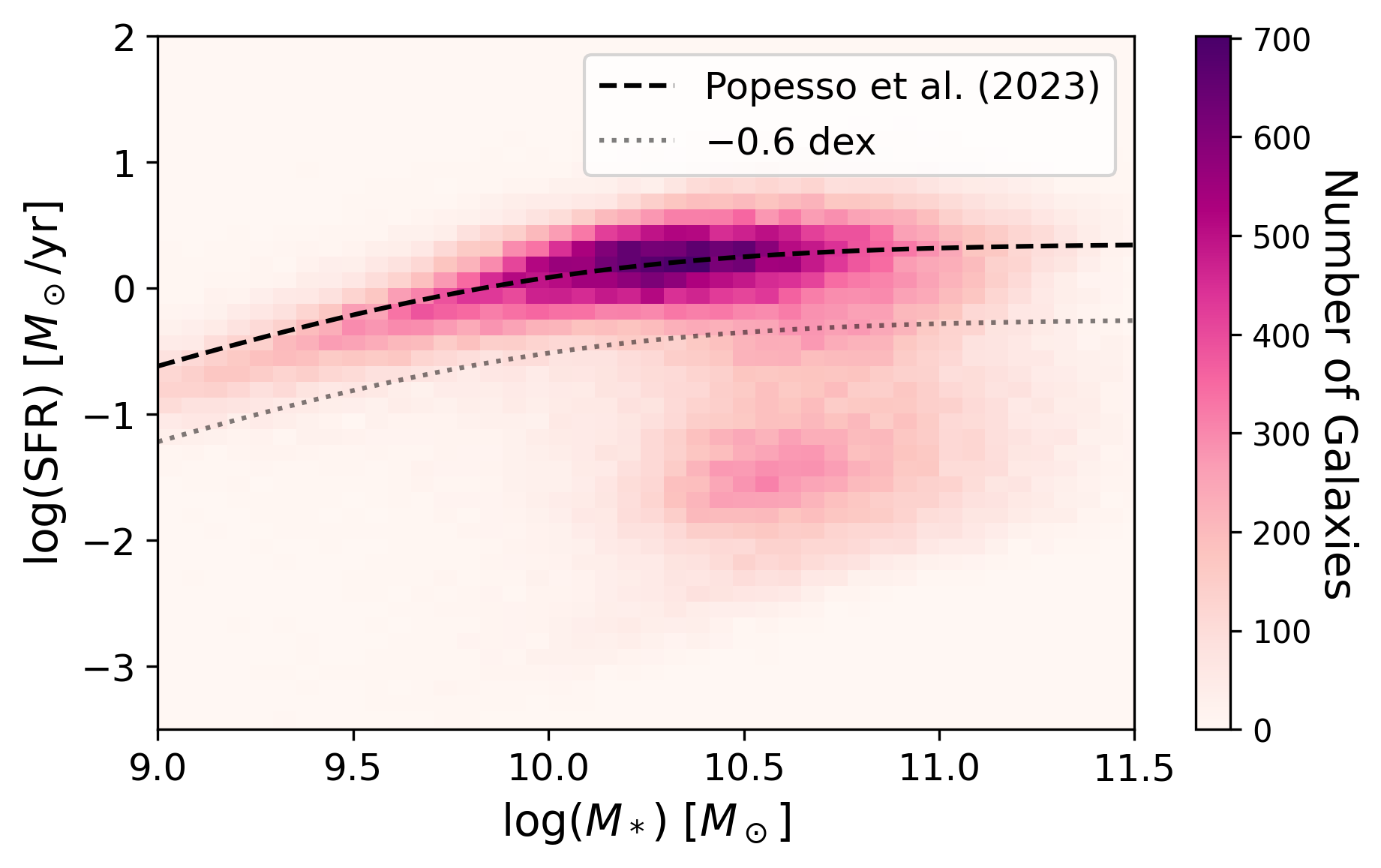}
    \caption{SFR vs stellar mass for the parent sample (138956 group, cluster, and field galaxies). The SFMS as defined by \citep{popesso_2023} is shown by the black dashed line, and our star-forming sample is selected to be all galaxies above the grey dotted line.}
    \label{figure_SFMS}
\end{figure}

We also restrict our sample to only include galaxies on the star-forming main sequence (SFMS), using the redshift-dependent SFMS defined by \cite{popesso_2023} at $z=0.067$ (the average redshift of our parent sample). We take all galaxies $0.6$dex below the line and above as star-forming. The distribution of our parent sample (all group, cluster, and field galaxies with $M_*\geq10^{9}M_\odot$ in both the \cite{yang_2007} and GSWLC-M2 \citep{salim_2016,salim_2018} catalogues) in stellar mass and SFR is shown in Figure \ref{figure_SFMS}, with the SFMS defined by \cite{popesso_2023} over-plotted.

\begin{figure}
    \centering
    \includegraphics[width=\linewidth]{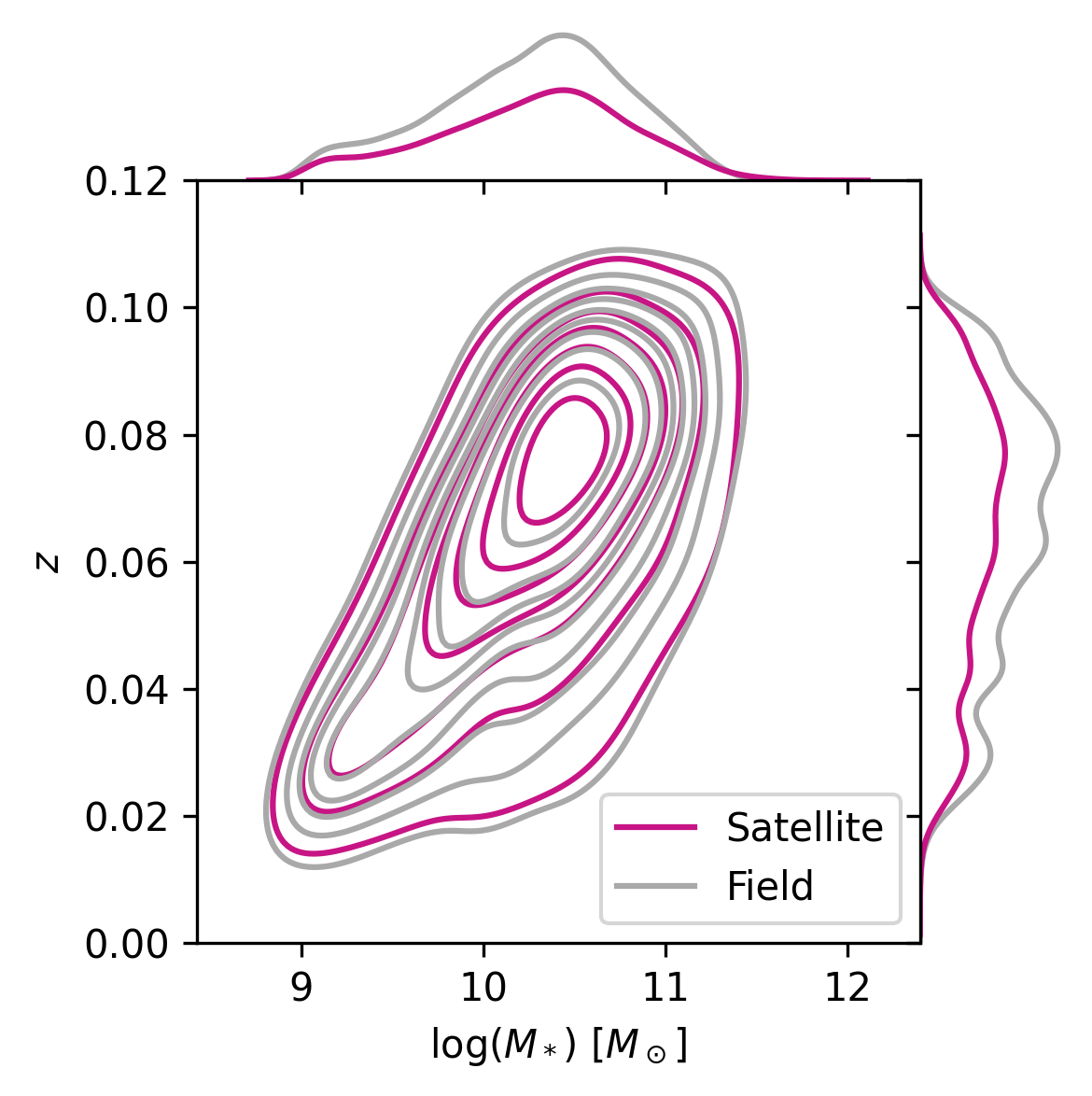}
    \caption{The distributions of satellite (magenta) and field (grey) samples in redshift-stellar mass space. Above and to the right of the main plot are one-dimensional kernel-density estimate (KDE) distributions of redshift and stellar mass respectively.}
    \label{figure_kde}
\end{figure}

As the measures we use depend on resolution and stellar mass (see Section \ref{section_properties}), we stellar mass and redshift-match the field sample to the satellite sample. To do this, we create 2D bins of width $0.1\text{dex}$ in stellar mass and $0.005$ in redshift. For each stellar mass and redshift bin, we randomly select $\sim2$ field galaxies for each satellite galaxy. If within a bin there are not enough field galaxies to correspond to twice the number of satellite galaxies, we simply take all the galaxies available in that bin. Figure \ref{figure_kde} shows the distribution of each sample in redshift-stellar mass space.

Mass matching is crucial for our analysis as the effectiveness of ram pressure has a strong dependence on galaxy stellar mass: higher mass galaxies can hold onto their star-forming gas more readily than lower mass galaxies \citep{gunn_1972,hester_2006,steyrleithner_2020,boselli_2022}. We also found that many of our star formation asymmetry measures (as described in Section \ref{section_SFasyms}) have a strong mass dependence, with asymmetry increasing as stellar mass decreases. 

\subsection{UNIONS imaging}
\label{section_CFISimaging}

The UNIONS collaboration is a partnership between wide-field imaging surveys of the northern hemisphere, consisting of the Canada-France Imaging Survey \citep[CFIS, deep $u$ and $r$-band;][]{ibata_2017}, Pan-STARRS (deep $i$ and moderate-deep $z$-band), Wide Imaging with Subaru HSC of the Euclid Sky (WISHES, deep $z$-band imaging), and the Waterloo-Hawaii IfA $g$-band Survey (WHIGS, $g$-band). These surveys were designed, in part, to secure optical imaging to complement the Euclid space mission (\citealt{mellier_2024}; Gwyn et al.~2024, in preparation), while UNIONS is a separate collaboration aimed at maximizing the science return of these large and deep surveys of the northern skies.

CFIS consists of high-quality $u$ and $r$-band imaging (seeing $\sim1"$ and $\sim0.7"$ respectively) covering $\sim5000$ square degrees of the sky. The CFIS/UNIONS $u$-band reaches a depth of 23.6 magnitudes and the $r$-band reaches a depth of 24.1 magnitudes for point sources measured in an aperture of 2 arcseconds at a signal-to-noise ratio of 5. We use the $u$-band to trace active star formation, noting that the $u$-band is affected by dust extinction. However, the $u$-band is less affected by dust extinction than the UV, with an average dust attenuation factor of $A_u\sim1.0\text{mag}$ at $z=0$ \citep{prescott_2009}, increasing with redshift and SFR \citep{hopkins_2003,prescott_2009,figueira_2022}. We use the $r$-band to trace the older stellar population for the galaxies in our sample.

We make cutouts in both bands using the Canadian Astronomy Data Centre UNIONS cutouts service, choosing cutouts to be squares with side length $10r_e$, where $r_e$ is the SDSS $r$-band effective radius from Table 3 of the \cite{simard_2011} morphology catalogue. We exclude galaxies with signal-to-noise ratios less than 100 in the $u$-band and apply the same cut on the $r$-band when measuring $u-r$ colour.

We measure asymmetries in star formation in the UNIONS $u$-band (Section \ref{section_SFasyms}) within apertures defined by $u$-band Sérsic profiles, fit using the astronomical image fitting program \texttt{imfit} \citep{erwin_2015} with the \cite{simard_2011} SDSS $r$-band pure Sérsic parameters as initial guesses for the iterative process. Our results are consistent with apertures defined using the $r$-band imaging, and since the current sky coverage of the UNIONS $u$-band is $\sim2\times$ that of the $r$-band we proceed with the $u$-band apertures. We exclude galaxies with inclination angles $i>75\degree$ to exclude galaxies at high inclination where measurements of asymmetry are unreliable \citep{giese_2016}. Our results are not dependent on the choice of inclination angle cutoff.

We restrict our analysis on the UNIONS imaging to elliptical apertures centered on the galaxy coordinates with ellipticities and position angles defined by the Sérsic parameters, with semi-major axes of two effective radii ($2r_e$). To make background measurements, we use an elliptical annulus with an inner radius of $3r_e$ and an outer radius of $5r_e$ with the same parameters. We note that our results are independent of the choice of aperture size. In addition, we tested masking the inner regions of galaxies to increase sensitivity to effects on the leading and trailing edges but found no significant changes in our results.

With the selection of the data described above, we are left with a satellite sample of 5277 galaxies and a stellar mass and redshift-matched field sample of 8360 galaxies. Stellar masses range from $10^{9.0}M_\odot<M_*<10^{11.7}M_\odot$, redshifts from $0.01<z<0.1$, and halo masses from $10^{13.0}M_\odot<M_h<10^{15.1}M_\odot$.

\section{Methods}
\label{section_3}

\subsection{Projected phase space}
\label{section_phasespace}
Different regions of projected phase space (PPS) are occupied by galaxies with specific times since infall \citep[e.g.][]{mahajan_2011,oman_2013,rhee_2017,pasquali_2019}, with galaxies falling into groups and clusters on a zig-zag trajectory through PPS on average (see Figure 4 in \cite{yoon_2017}). To assign a time since infall ($t_{inf}$) to the galaxies in our sample, we use the regions defined by \cite{rhee_2017} in PPS. To place the galaxies in PPS, we use the projected distance between the galaxy centre and the group/cluster centre, computed using the angular diameter distance and line-of-sight velocity relative to the group/cluster velocity. These values are normalized by the group/cluster virial radius and velocity dispersion respectively.

Using cosmological hydrodynamic N-body simulations, \cite{rhee_2017} divide PPS into five regions according to how long an average galaxy has been living in a group or cluster environment. Region A hosts the most first-infall galaxies (galaxies falling into the cluster for the first time, $t_{inf}=0$), region B hosts the highest fraction of recent infallers ($t_{inf}<3.63\text{Gyr}$), regions C and D host intermediate infallers ($3.63\text{Gyr}<t_{inf}<6.45\text{Gyr}$), and region E is dominated by ancient infallers ($t_{inf}>6.45\text{Gyr}$). However, we note that the regions are contaminated by galaxies with varying time since infall, especially region A which hosts a large fraction of interlopers. We combine regions C and D in our analysis as they host similar fractions of first, recent, intermediate, and ancient infallers. See Figure 6 in \cite{rhee_2017} for the fraction of each population in each region of phase space. 

\begin{figure}
    \centering
    \includegraphics[width=\linewidth]{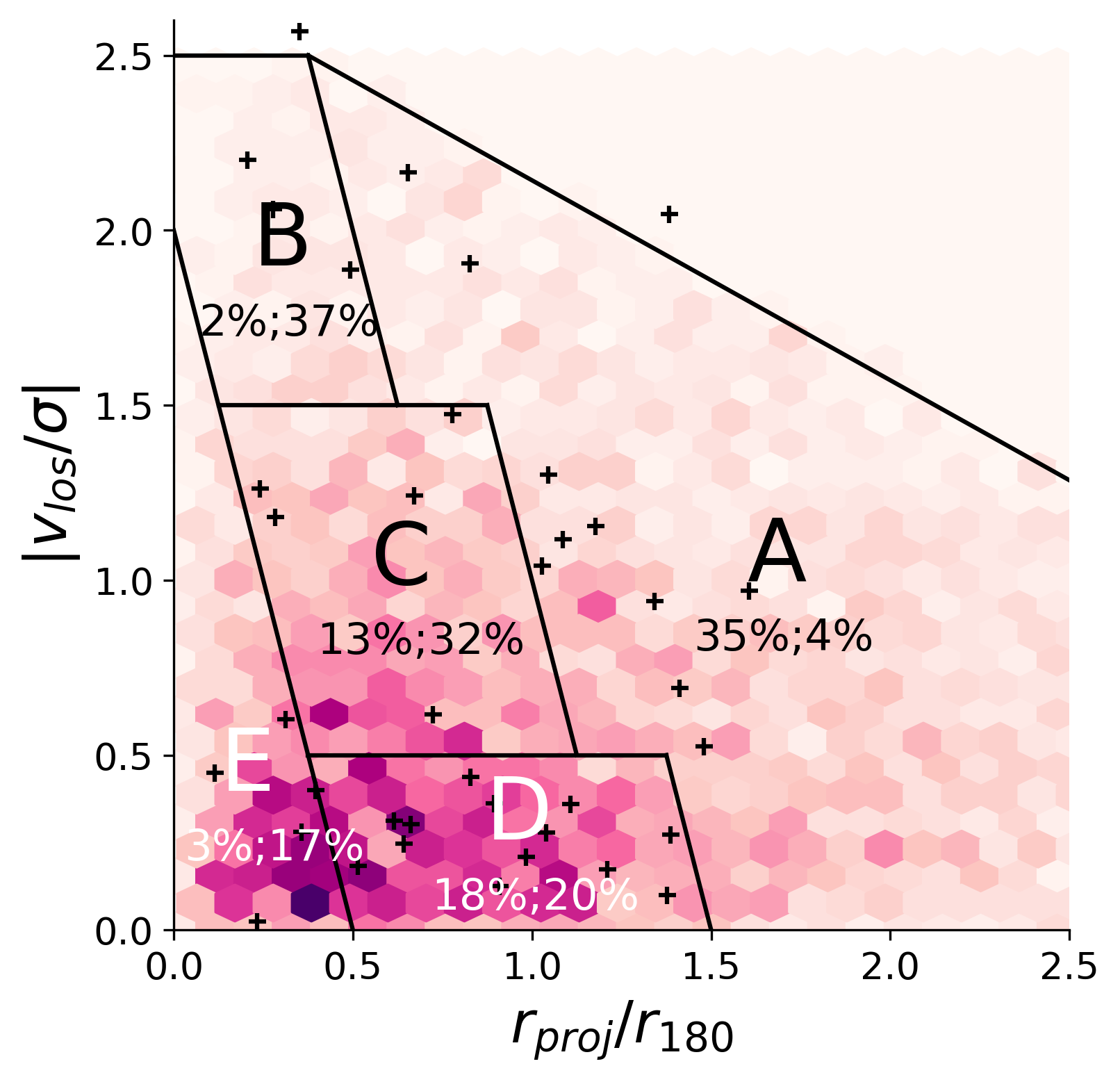}
    \caption{Density plot of the satellite samples in projected phase space. Overlaid are the phase space regions as defined by \protect\cite{rhee_2017}, where the average time since infall of each region increases from A to E. The positions of the Coma Cluster jellyfish galaxies are shown with black crosses. Below each label is the fraction of the first/recent infaller populations in that region of PPS according to \cite{rhee_2017}.}
    \label{figure_A}
\end{figure}

With this division of phase space, we can explore trends between star formation asymmetries and the amount of time a galaxy has spent in a dense environment. We would expect enhanced star formation to be strongest in galaxies in clusters that have infallen recently, as ram pressure would be strong, but would not yet have had enough time to strip the galaxy of its gas supply, i.e., galaxies in regions A and B. In Figure \ref{figure_A}, we show the distribution of our sample in PPS, with the \cite{rhee_2017} regions overlaid. Other properties such as stellar mass, SFR, and morphology of infalling galaxies have been shown to change as galaxies move through different regions of PPS, with galaxies becoming more passive on average as they move towards the virialized region of PPS \citep[e.g.][]{mahajan_2011,barsanti_2018,oxland_2024}.

\subsection{Star formation asymmetry measures}
\label{section_SFasyms}

\begin{figure*}
    \centering
    \includegraphics[width=\linewidth]{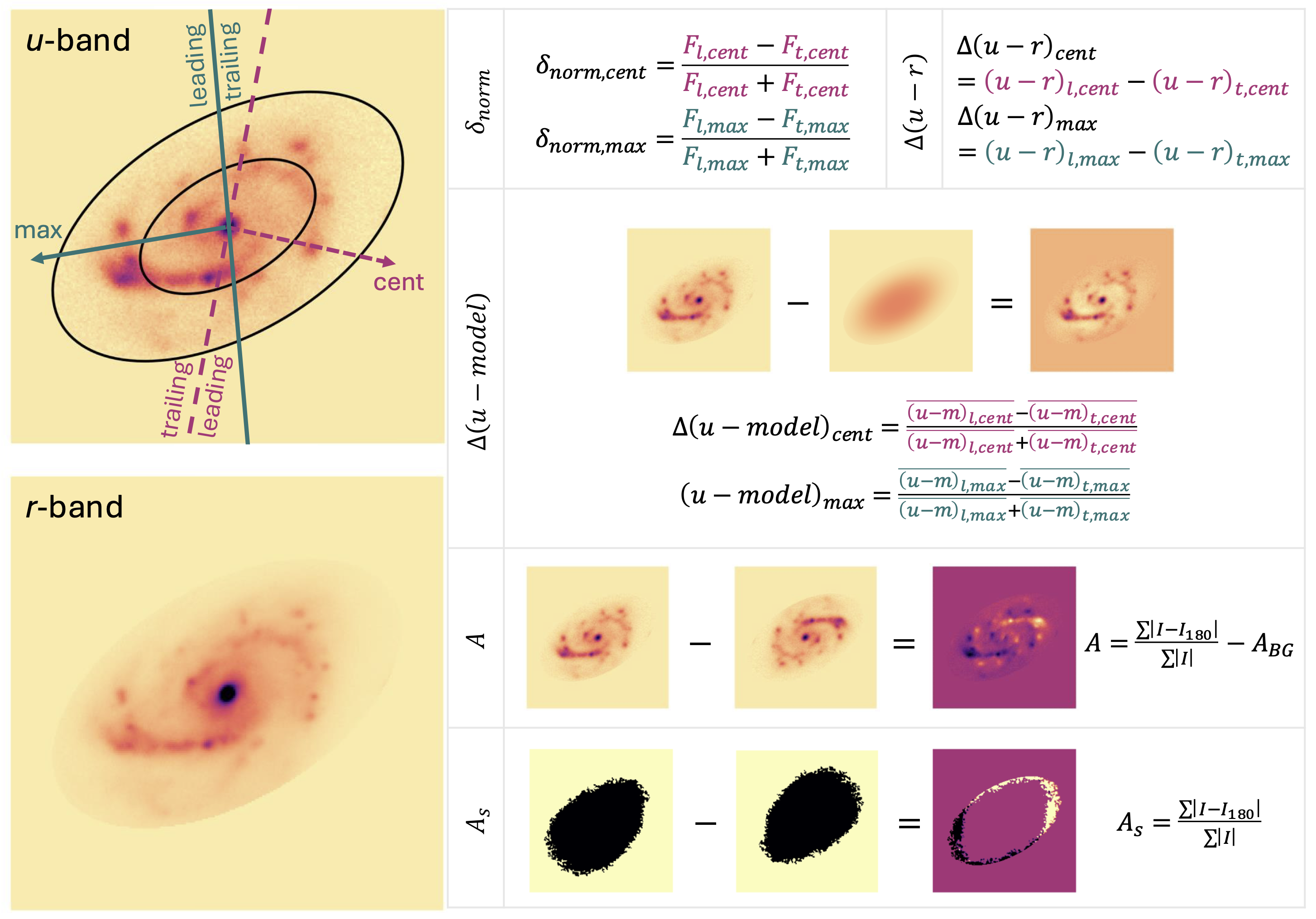}
    \caption{A schematic showing each of our star formation asymmetry measures for one galaxy in the satellite sample. The top left panel shows the galaxy in the UNIONS $u$-band. The outer black ellipse shows the $2r_e$ aperture in which we make our measurements, and the inner black ellipse shows a $1r_e$ aperture. The magenta dashed arrow points toward the group/cluster centre, and the pink dashed line shows the division between the assumed leading and trailing sides, with the leading side being the side facing the arrow. The green solid lines represent the same divisions, but for the direction which maximizes $\delta_{norm}$ (see equation (\ref{equation_5})). Note that the directions which maximize $\delta_{norm}$, $\Delta(u-model)$, and $\Delta(u-r)$ are typically the same for a given galaxy. The bottom left panel shows the galaxy in the UNIONS $r$-band. The equations in magenta correspond to the group/cluster centre division ($cent$), and the equations in green correspond to the maximized asymmetry division ($max$). In the $\Delta(u-model)$ panel, the left, centre, and right images show the $u$-band image, $u$-band Sérsic model, and the difference image respectively. In the $A$ panel, the left, centre, and right images show the $u$-band image, rotated $u$-band image, and difference image respectively. Finally, in the $A_s$ panel, the left, centre, and right images show the galaxy binary map, the rotated binary map, and the difference image respectively.}
    \label{figure_1}
\end{figure*}

We use five separate resolved star formation measures to quantify asymmetries in satellite galaxy star formation, illustrated in Figure \ref{figure_1}. Three of our measures are direction dependent, and directly compare the leading and trailing halves of the galaxy. Since we do not know the direction of motion we make the assumption that the galaxy is falling directly towards the group or cluster centre in projection. This has been shown to be a fair assumption for jellyfish galaxies falling into clusters \citep{roberts_2021a,roberts_2021b,roberts_2022}, where their true (projected) direction of motion is known through the orientation of their tail. This assumption may not work as well for group galaxies \citep{roberts_2021b} as the timescale for stripping is longer in groups, where the ICM is less dense and the galaxies are moving more slowly \citep{oman_2021,smith_2022}, and thus jellyfish in groups may be backsplashing when their tail forms \citep{roberts_2021b}. Measurements made using this cut are labelled with a subscript of `$cent$'. For the field galaxies in our sample, we randomly choose five angles between $0$ and $360\degree$ to cut the galaxies in half five ways, allowing us to increase the size of the field sample.

Given the directional uncertainty, we also identify the cut which maximizes each measure, as shown in Figure \ref{figure_1}, for a total of eight measurements of star formation asymmetry. \cite{troncosoiribarren_2020} showed that the cut which maximizes star formation asymmetry in 3D is well correlated with the direction of motion for infalling galaxies in the EAGLE simulations \citep{schaye_2015}. They compared a radial cut, assuming the galaxy is falling directly towards the cluster centre, to a maximizing cut and found that the maximizing cut is much more effective at detecting star formation asymmetries caused by ram pressure effects. They suggest this is the best observational cut to study the effects of ram pressure on infalling galaxies in large samples. These measures are labelled with a subscript of `$max$'.

In the following subsections, we describe the methods used, and present our findings using these methods in Section \ref{section_4}. Figure \ref{figure_1} shows a visualization of each of the methods. We generalize the definition of the phrases `leading' and `trailing' here, which refer to the halves in both the centre/$cent$ method (which defines the `leading' side as that facing the group/cluster centre in projection) and the maximized/$max$ method (which defines the `leading' side as the choice that maximizes that star formation asymmetry).

\subsubsection{Normalized flux difference}
For this method, we divide the galaxy in half to define leading and trailing sides. Then, we simply sum the flux on the leading and trailing sides, and take the normalized difference:

\begin{equation}
    \label{equation_4}
    \delta_{norm,cent}=\frac{F_{\text{leading,cent}}-F_{\text{trailing,cent}}}{F_{\text{leading,cent}}+F_{\text{trailing,cent}}}
\end{equation}

\noindent or

\begin{equation}
    \label{equation_5}
    \delta_{norm,max}=\frac{F_{\text{leading,max}}-F_{\text{trailing,max}}}{F_{\text{leading,max}}+F_{\text{trailing,max}}}
\end{equation}

Thus, a value of 1 means all the flux is on the leading side, a value of 0 means there is equal flux on both sides and a value of -1 means all the flux is on the trailing side. As $u$-band flux is directly correlated with SFR \citep[e.g.][]{moustakas_2006}, we can take this to be the normalized SFR difference between sides without the extra step of converting the flux. A similar measure was used by \cite{troncosoiribarren_2020}, where they found that infalling galaxies in the EAGLE simulation have higher values of $\delta_{norm}$ as compared to the field.

\subsubsection{Deviation from the model}
\label{subsection_u-model}
This method compares the average deviation from the galaxy from its $u$-band Sérsic model $m$ on the leading and trailing halves. We use a similar equation to equations (\ref{equation_4}) and (\ref{equation_5}), but instead we take the mean of the deviations:

\begin{equation}
    \Delta(u-m)_{cent}=\frac{\overline{(u-m)_{\text{leading,cent}}}-\overline{(u-m)_{\text{trailing,cent}}}}{\overline{(u-m)_{\text{leading,cent}}}+\overline{(u-m)_{\text{trailing,cent}}}}
\end{equation}

\noindent or

\begin{equation}
    \Delta(u-m)_{max}=\frac{\overline{(u-m)_{\text{leading,max}}}-\overline{(u-m)_{\text{trailing,max}}}}{\overline{(u-m)_{\text{leading,max}}}+\overline{(u-m)_{\text{trailing,max}}}}
\end{equation}

\noindent where a value of 0 indicates the star formation is not asymmetric. A positive value indicates the star formation is above what is expected on the leading side, or that the star formation is lower than expected on the trailing side, with the reverse being true for a negative value.

\subsubsection{Difference in u-r colour}
The final direction-dependent measure takes the difference in $u-r$ colour between halves, as $u-r$ colour traces specific star formation rate (sSFR) \citep{kauffmann_2003,salim_2007}. For this method, we take the subset of galaxies in our sample with both $u$ and $r$-band imaging, resulting in a satellite sample of 2588 galaxies and a field sample of 3853 galaxies. We calculate the $u-r$ colours for each half of the galaxies and take the difference:

\begin{equation}
    \Delta(u-r)_{cent}=(u-r)_{\text{leading,cent}}-(u-r)_{\text{trailing,cent}}
\end{equation}

\noindent or

\begin{equation}
    \Delta(u-r)_{max}=(u-r)_{\text{leading,max}}-(u-r)_{\text{trailing,max}}
\end{equation}

A negative difference in $\Delta(u-r)$ would indicate higher sSFR on the leading side. \cite{rodriguez_2020} used a similar method in a study of $g-r$ colour on the leading and trailing halves of infalling galaxies in SDSS, assuming galaxies are falling directly into the group/cluster centre in projection. They find that the regions on the leading halves of infalling galaxies have slightly bluer $g-r$ colours for galaxies beyond $0.75r_{180}$ with their semi-major axes perpendicular to the projected group centre direction.

\subsubsection{Asymmetry}
\label{subsection_casasymmetry}

We use galaxy global asymmetry $A$ from the CAS (Concentration, Asymmetry, Smoothness) system \citep{conselice_2003,conselice_2014}. We adopt a similar definition of $A$ to that used by \cite{lotz_2004}. We first mask sources that are not a part of the galaxy using the functions \texttt{detect\textunderscore threshold} and \texttt{detect\textunderscore sources} from the \texttt{photutils} package \citep{bradley_2024} applied to a smoothed version of the image, where we require that the pixels contained in the source must be $3\sigma$ above the background noise, and the source must contain at least $40$ connected pixels. We then apply another mask which removes all pixels above and below $6\sigma$ within the galaxy, excluding the bright centre with an elliptical annulus with inner and outer radii $1$ and $2r_e$ and $2\sigma$ within the background (elliptical annulus with inner and outer radii $3$ and $5r_e$). We finally rotate the image $180\degree$, and compute $A$ using the equation:

\begin{equation}
    A = \frac{\sum|I-I_{180}|}{\sum|I|} - \frac{n_{pix,I}\sum|B-B_{180}|}{n_{pix,B}\sum|I|}
\end{equation}

\noindent where $I$ and $B$ are the galaxy and background components of the image respectively, a subscript of $180$ indicates that component has been rotated $180\degree$, and the sums are over all unmasked pixels for each component. The background term is adjusted to contribute the same number of pixels as the galaxy component by multiplying by a factor of $n_{pix,I}/n_{pix,B}$. We note that $A$ is strongly affected by noise, so it is critical to carefully account for the background of the image.

\subsubsection{Shape Asymmetry}

\cite{pawlik_2016} define shape asymmetry $A_s$, a morphological parameter which is similar to $A$ but instead computes the asymmetry value from a binary map weighing all detected pixels equally. We follow the method outlined in \cite{pawlik_2016}, using the same mask as in Section \ref{subsection_casasymmetry}, but clipping at $3\sigma$ rather than $6\sigma$ to follow \cite{pawlik_2016}. To make faint features more easily detectable, the image is first smoothed using a box filter. The binary map is then created from the smoothed version of the image by accepting 8-connected pixels as a part of the galaxy. A similar equation to that of CAS asymmetry is used to calculate shape asymmetry:
\begin{equation}
    A_s = \frac{\sum |I-I_{180}|}{\sum I}
\end{equation}

\noindent where $I$ is now the binary map of the galaxy. We refer the reader to \cite{pawlik_2016} for an in-depth description of the method. This method allows for all components of the galaxy to be weighted equally instead of by brightness due to the use of a binary map, which is beneficial for identifying faint outer features such as stripped tails. We note that like $A$, $A_s$ is also strongly affected by noise. We also tested an alternative to $A$ and $A_s$, root-mean-squared asymmetry \citep{sazonova_2024}, which is less sensitive to noise. However, we found that this measure still has a noise dependence in our data and we choose to not include the results here. $A_s$ was used by \cite{roberts_2020} in a study of local cluster galaxies where it was found that jellyfish galaxies tend to have higher $A_s$ than both star-forming and quenched non-stripping cluster galaxies.

\section{Results}
\label{section_4}

\subsection{Test case on Coma Cluster ram pressure stripped sample}
\label{section_comacluster}
To ensure the star formation asymmetry measures described in Section \ref{section_3} can detect signatures of ram pressure stripping, we check that a known ram pressure-stripped sample of galaxies is distinct from a sample of normal star-forming galaxies in these measures. We use two samples of jellyfish galaxies in the Coma Cluster to compare against the normal star-forming Coma cluster galaxies in our sample.

\cite{roberts_2020} present a sample of 41 jellyfish galaxies in the Coma Cluster visually identified in CFHT three-colour $ugi$ imaging. \cite{roberts_2021a} present a sample of 95 jellyfish galaxies in low redshift ($z<0.05$) clusters with extended radio tails in LOFAR Two-metre Sky Survey (LoTSS; \cite{shimwell_2017,shimwell_2019}) 144 MHz images. Of the 95 jellyfish galaxies, 29 are within the Coma Cluster, with significant overlap with the \cite{roberts_2020} sample. We show the distribution of Coma Cluster jellyfish galaxies in PPS in Figure \ref{figure_A}.

We use both samples of Coma Cluster jellyfish galaxies in our test case as the \cite{roberts_2020} sample is likely more complete and goes to lower stellar mass, while the \cite{roberts_2021a} sample is likely more pure but has a higher detection threshold and thus excludes lower mass and lower SFR galaxies from the sample. The \cite{roberts_2021a} sample also detects clear tails, while the \cite{roberts_2020} sample does not. We follow the steps in Section \ref{section_3} to compute the star formation asymmetries of these galaxies; however as very few have UNIONS $r$-band imaging we do not compute the $\Delta(u-r)$ measures.

Matching the selection and cuts on our total sample, we obtain a sample of 32 jellyfish galaxies from \cite{roberts_2020}, 25 jellyfish galaxies from \cite{roberts_2021a}, and 112 control normal star-forming Coma Cluster galaxies taken from our total cluster sample (from which both samples of jellyfish galaxies are removed). For each jellyfish sample, we create a 1:1 mass-matched normal Coma Cluster sample within bins of $0.1\text{dex}$ in stellar mass to remove any mass-dependence.

\begin{figure*}
    \centering
    \includegraphics[width=\linewidth]{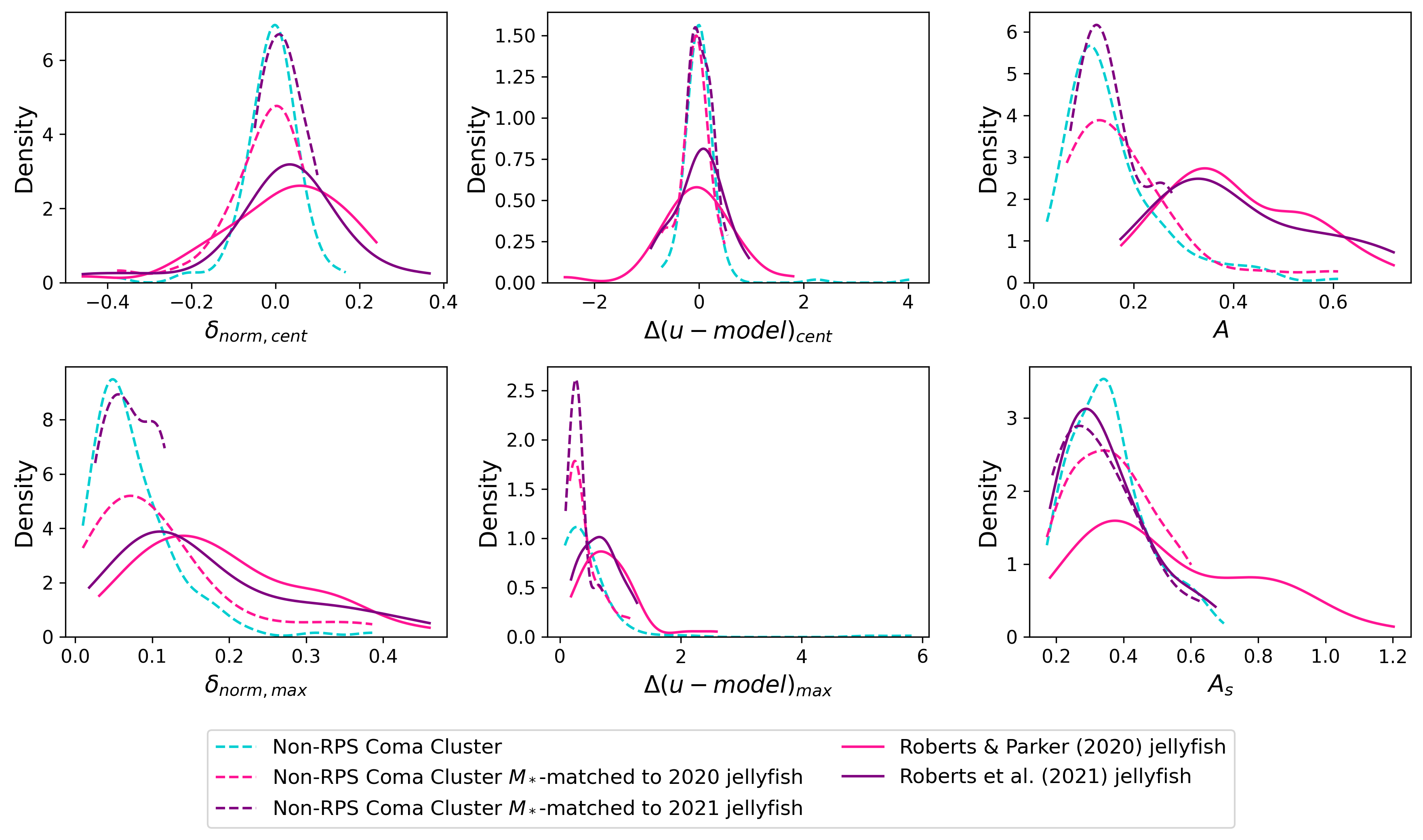}
    \caption{KDE distributions of star formation asymmetry measures for the Coma Cluster. The dashed teal lines correspond to the full normal star-forming Coma Cluster sample. The solid pink and purple lines correspond to the \protect\cite{roberts_2020} and \protect\cite{roberts_2021a} jellyfish samples. The dashed pink and purple lines correspond to the normal Coma Cluster sample stellar mass matched to the \protect\cite{roberts_2020} and \protect\cite{roberts_2021a} jellyfish samples.}
    \label{figure_2}
\end{figure*}

In Figure \ref{figure_2} we show the distributions of our star formation asymmetry measures for each sample. In most measures, we can see that the jellyfish samples are skewed to higher values than their mass-matched Coma Cluster samples. We compare the similarity between jellyfish and normal Coma Cluster galaxy distributions by performing a two-sample Anderson-Darling (AD) test using \texttt{scipy.stats.anderson\textunderscore ksamp}, and consider two samples to be distinct if their AD test p-value is $<0.05$. All AD test p-values are $<0.05$ except for the \cite{roberts_2021a} sample in $\delta_{norm,cent}$ and $A_s$ and both samples in $\Delta(u-model)_{cent}$. We also test degrading our images to simulate higher redshift imaging up to $z=0.1$ and find that these results hold even at higher redshift. We conclude that our methods are able to distinguish the jellyfish population from the Coma Cluster as a whole even at $z=0.1$, and can apply our method to the larger sample of low redshift galaxies described in Section \ref{section_2}.

In the remainder of the paper, we focus on measures which had distinct distributions for the jellyfish sample and discard all methods which assume the galaxies are falling directly into the group/cluster centre in projection, i.e., $\delta_{norm,cent}$, $\Delta(u-model)_{cent}$, and $\Delta(u-r)_{cent}$, as the methods adopting this assumption performed most poorly in the Coma Cluster test case. This is further discussed in Section \ref{section_discussion_coma}. We provide readers with a data table including galaxy properties ($M_*$, SFR, position angle $\theta$, ellipticity $e$, effective radius $r_e$, SNR, and inclination angle $i$), environmental properties ($M_h$, PPS region), and asymmetry measurements for our satellite and field samples as supplementary online material, described in Appendix \ref{appendix}.

\subsection{Star formation asymmetry distributions}
\label{section_asymmetry_distributions}

\begin{figure*}
    \centering
    \includegraphics[width=\linewidth]{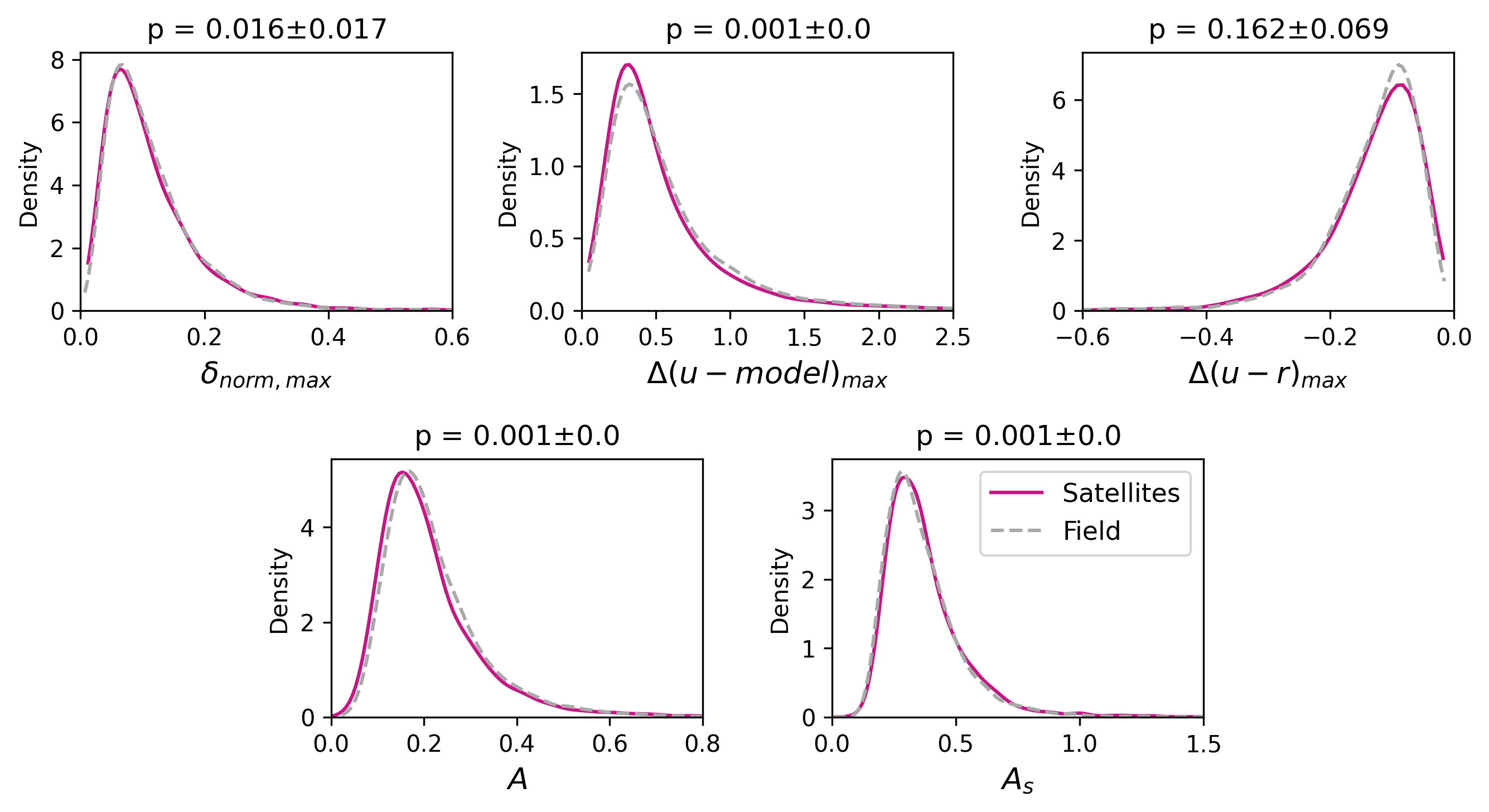}
    \caption{KDE distributions of star formation asymmetry measures for the full satellite (solid magenta lines) and field (dashed grey lines) samples. Each panel shows a different star formation asymmetry measure, with the AD test p-values and their $1\sigma$ errors comparing the two populations given above each panel.}
    \label{figure_3}
\end{figure*}

We first compare the full star formation asymmetry distributions of our satellite sample to those of the control field sample as shown in Figure \ref{figure_3}. As the field sample is $\sim$twice as large as the satellite sample, we number match the field sample to the satellite sample to ensure a fair comparison between populations when running an AD test. We construct 100 number-matched field samples by randomly drawing galaxies from the full field sample and run the AD test for each subsample. This allows us to compute a mean p-value with uncertainty. In Figure \ref{figure_3}, the p-values and their $1\sigma$ errors are given for each star formation asymmetry measure. Visually, we see that for each measure the distributions for the satellite sample and the field sample are extremely similar. The AD test picks up on subtle differences in $A$, $A_s$, $\delta_{norm,max}$, and $\Delta(u-model)_{max}$, as indicated by low p-values, which appear to be caused by slightly higher values in the field sample. The AD test is able to pick up on tiny distinctions between large datasets, as illustrated here with 5277 datapoints in each sample. We note that all our results are consistent if using either the Kolmogorov–Smirnov or the Anderson-Darling test.

\begin{figure*}
    \centering
    \includegraphics[width=\linewidth]{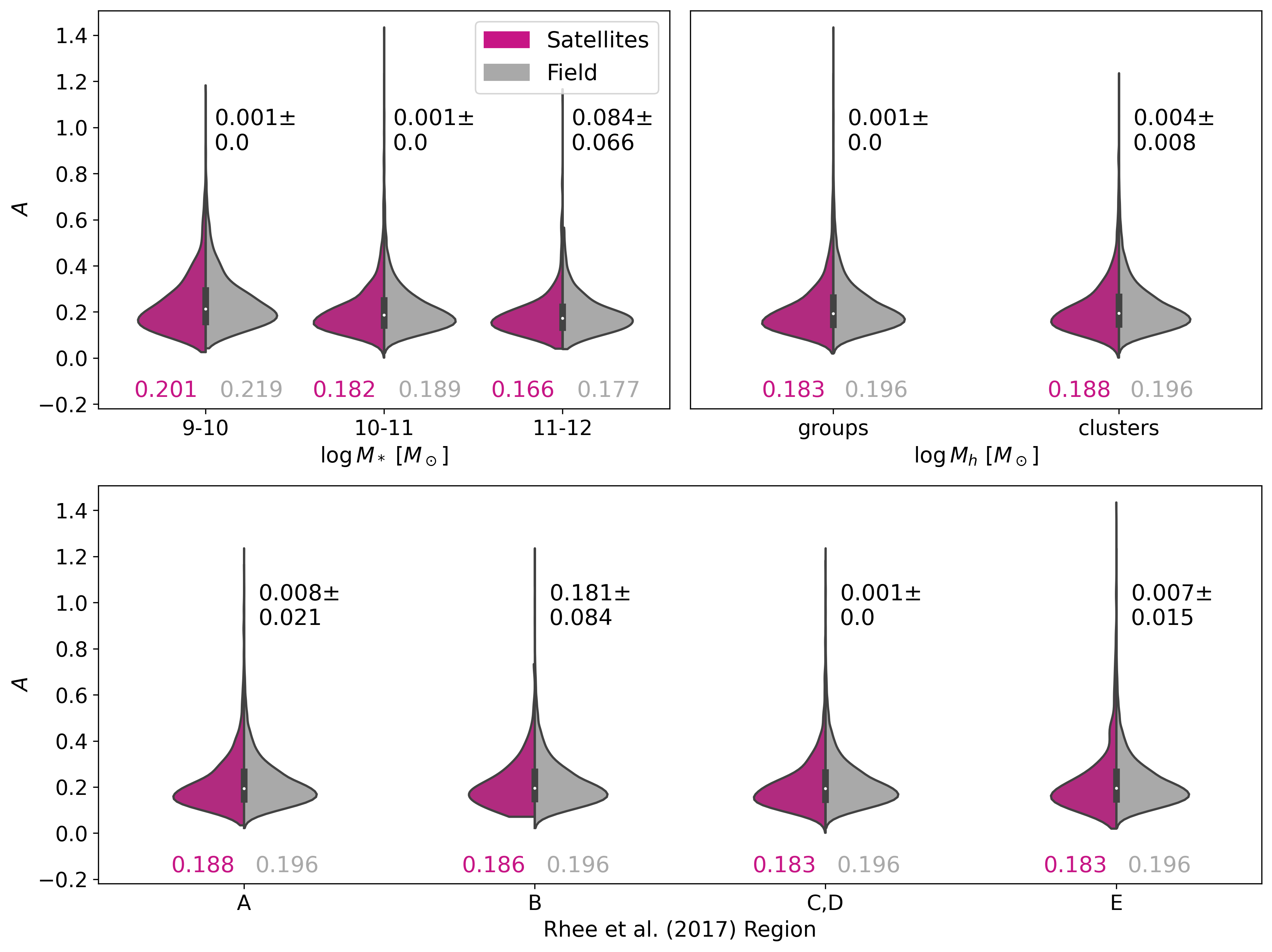}
    \caption{Violin plots showing the distribution of $A$ binned by different properties. Each violin (created using \texttt{seaborn.violinplot} \protect\citep{waskom_2021}) is a KDE distribution with an inner box plot showing the extent of the distribution excluding outliers (thin black lines), quartile range (thick black lines), and median of the distribution (white point). The satellite sample is shown in magenta, and the field sample is shown in grey. Note that the full field distribution is shown in each violin in the upper right and bottom panels, as $M_h$ and \cite{rhee_2017} regions are not relevant for field galaxies. Galaxies are binned by stellar mass, host halo mass, and time since infall in the top right, top left, and bottom panels respectively. The coloured values below each half-violin give the mean of the distribution, and the values to the right of each violin give the mean p-values and their $1\sigma$ errors comparing the satellite sample to the field sample. The complete figure set including the (4 other asymmetry measures) is available in the online journal.}
    \label{figure_4}
\end{figure*}

We further split the sample by galaxy and environmental properties to search for differences in each star formation asymmetry distribution. We divide the sample into bins of stellar mass, halo mass, and \cite{rhee_2017} PPS region, as shown for asymmetry measure $A$ in Figure \ref{figure_4}. We can divide the field sample by stellar mass, but as they have no host halo mass or time since infall, we show the full field distribution. Using the same method to number-match the satellite and field samples, we obtain p-values and their errors for each bin. We show this plot for $A$ only, in which we see that the satellite sample is distinct from the field for galaxies with stellar masses $10^9M_{\odot}\leq M_*<10^{11}M_\odot$, galaxies in groups or clusters, and galaxies with early (region A), intermediate (regions C and D), or late (region E) time since infall. We see similar results in all other measures, where there are no visually distinct populations, despite some comparisons having small p-values. Any differences seem to be caused by higher values in field galaxies, and we stress that these differences are small and the overall star formation asymmetry distributions are remarkably similar between field and group/cluster environments. Equivalent versions of Figure \ref{figure_4} for the other star formation asymmetry measures can be found in the supplementary material.

\subsection{Subsample of galaxies likely to be experiencing RPS}

\begin{figure*}
    \centering
    \includegraphics[width=\linewidth]{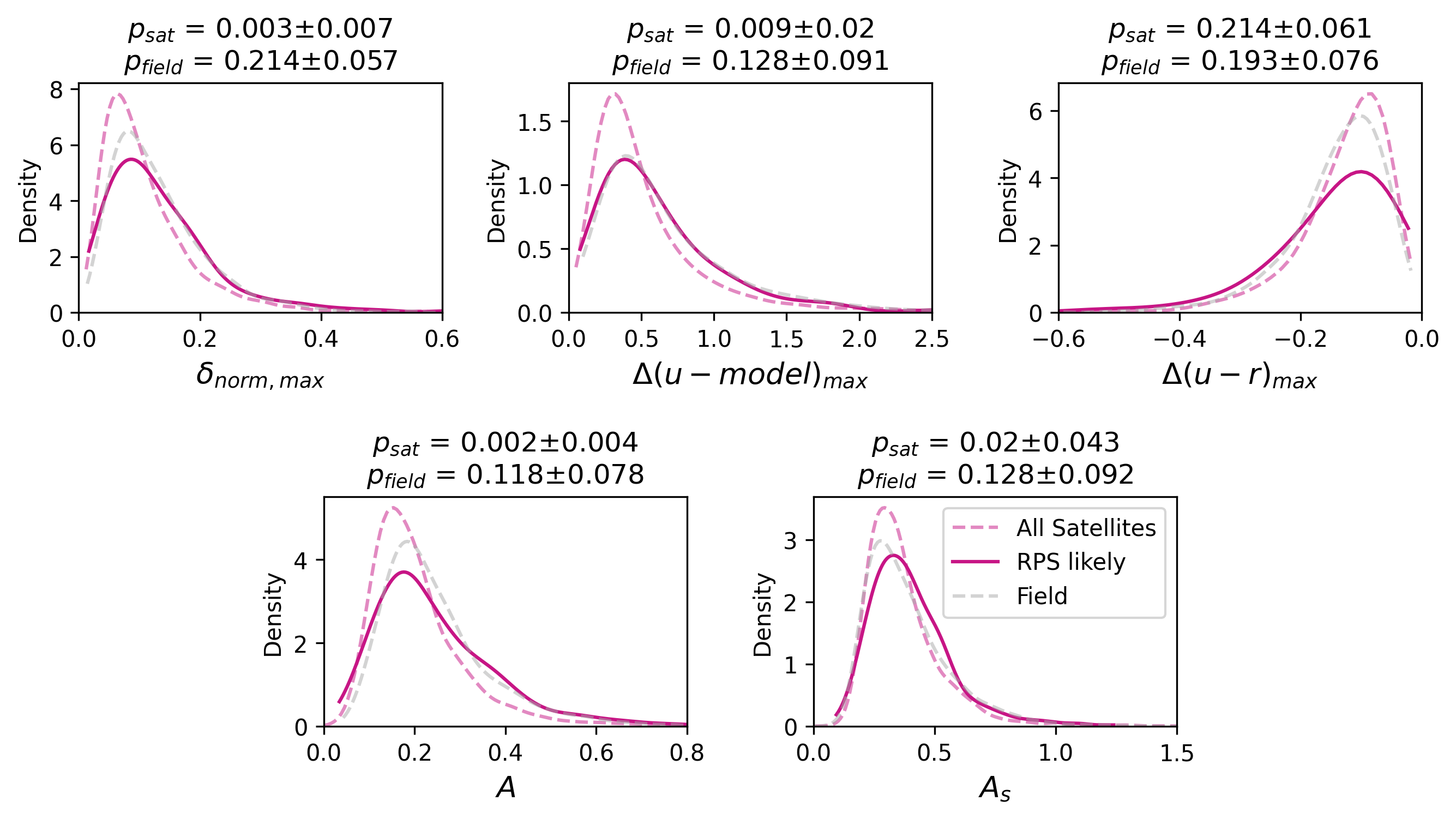}
    \caption{KDE distributions of star formation asymmetry measures for the subsample likely experiencing RPS (solid magenta lines), the full satellite sample (dashed magenta lines), and the low stellar mass field sample (dashed grey lines). Each panel shows a different star formation asymmetry measure, with AD test p-values and their $1\sigma$ errors comparing the two satellite samples ($p_{sat}$) and the RPS-likely and the low stellar mass field sample ($p_{field}$) given above each panel.}
    \label{figure_5}
\end{figure*}

Next, we investigate if the galaxies which we would most expect to be experiencing strong ram pressure have higher star formation asymmetries than other satellite galaxies. We create a sample of galaxies likely experiencing strong ram pressure prior to stripping by keeping only galaxies with low stellar mass ($M_*<10^{10}M_\odot$), high host halo mass (clusters: $M_h\geq 10^{14}M_\odot$), and low time since infall (\cite{rhee_2017} regions A and B). We obtain a subsample of 347 galaxies which are likely to be experiencing strong ram pressure. For the $\Delta(u-r)$ measures, the subsample is only 159 galaxies. We apply the same stellar mass cut on the field sample, obtaining a sample of 2445 field galaxies. In Figure \ref{figure_5}, we show the star formation asymmetries distributions of this sample as compared to the full satellite sample (with the subsample removed) and the stellar mass-matched field sample.

As in Figure \ref{figure_3}, we give AD test p-values comparing 100 number-matched samples drawn from the full satellite sample above each panel. By visual inspection, we see that both the satellite and field subsamples are skewed to slightly higher values in most measures, but again any differences are very small.

\subsection{Properties of galaxies with the strongest star formation asymmetries}
\label{section_properties}

\begin{figure}
    \centering
    \includegraphics[width=\linewidth]{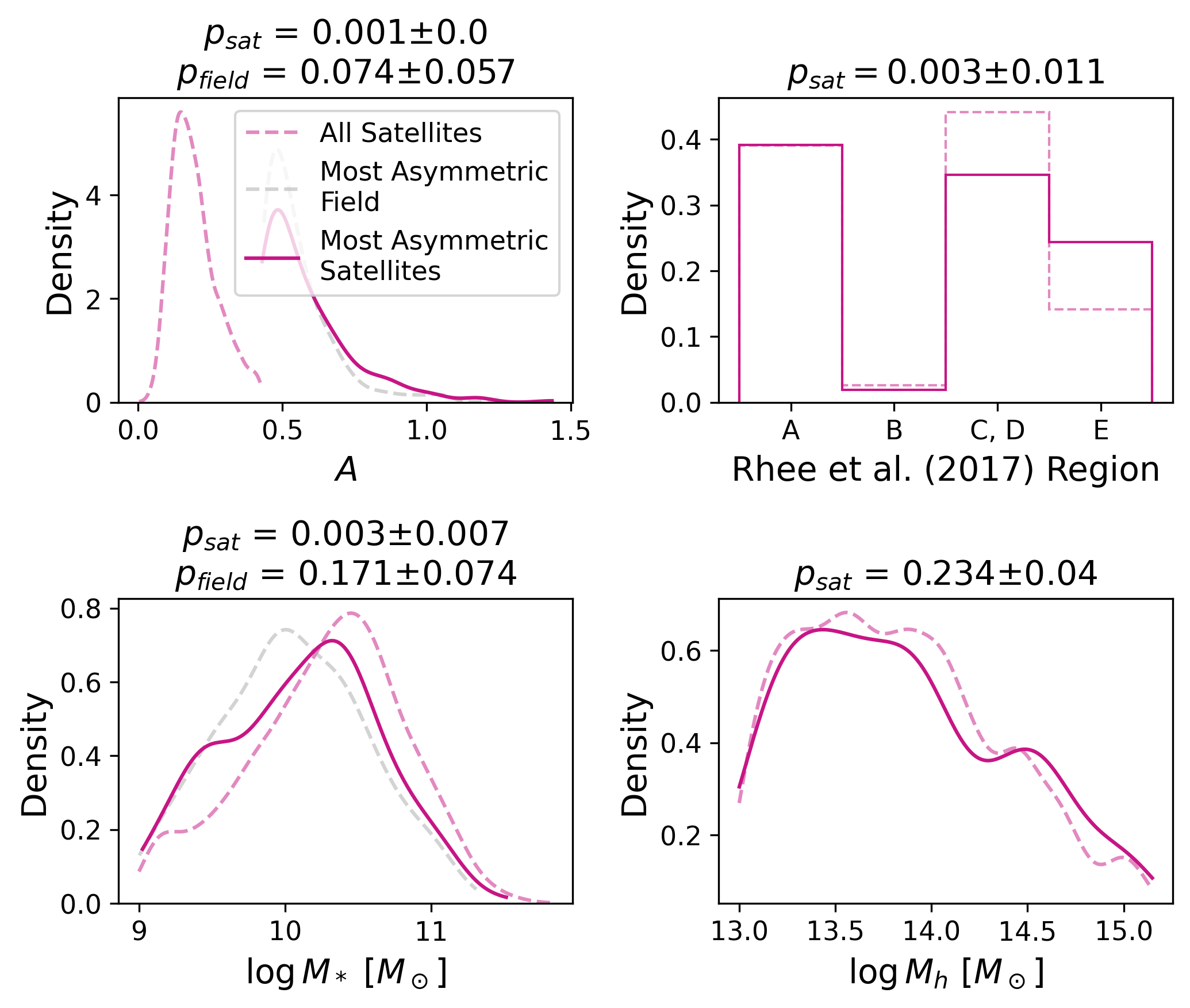}
    \caption{KDE distributions of asymmetry, stellar mass, and halo mass, and a histogram of \protect\cite{rhee_2017}. Distributions are shown for the top $5\%$ most asymmetric satellite galaxies in $A$ (solid magenta lines), the total satellite sample (dashed magenta lines), and the top $5\%$ most asymmetric field galaxies in $A$ (dashed grey lines). p-values and their $1\sigma$ errors comparing the two satellite samples ($p_{sat}$) and the highly asymmetric satellite and field samples ($p_{field}$) are given above each panel. p-values are computed using the AD test for continuous distributions and the chi-square test for discrete distributions. The complete figure set including the (4 other asymmetry measures) is available in the online journal.}
    \label{figure_6}
\end{figure}

As a final test, we can further subdivide our satellite sample by isolating the most asymmetric galaxies in the sample. First, we look at each measure individually and create a subsample of the top $5\%$ most-asymmetric galaxies in each measure (264 galaxies). We then compare the distributions of the properties of the galaxies in the most asymmetric sample and the full satellite sample. To control for stellar mass effects, we also compare to the top $5\%$ most-asymmetric field galaxies in each measure (418 galaxies). In Figure \ref{figure_6}, we show distributions of stellar mass, halo mass, and \cite{rhee_2017} region for galaxies in the top $5\%$ of the asymmetry measure $A$.

As before, we give the p-values comparing 100 number matched samples above each panel. For continuous distributions ($A$, $M_*$, $M_{h}$) we use the AD test as before, and the chi-square test (\texttt{scipy.stats.chisquare}) to compare the discrete \citep{rhee_2017} PPS region distributions. In most cases, the p-values comparing the properties of the full sample and the subsamples are greater than $0.05$ within $1\sigma$, except for the star formation asymmetry measures themselves, which is a result of the construction of our sample. By eye, we see that the most asymmetric sample is skewed to low stellar masses in all star formation asymmetry measures, but this is only statistically significant for the galaxies with high $A$. We also see that $A$ is skewed towards ancient infallers (region E), but this is not seen in any other measures. No other statistically significant differences between the satellite samples are observed, and the field is only statistically indistinguishable in stellar mass for $A_s$ where the distribution is still visually very similar. Equivalent versions of Figure \ref{figure_6} for the other star formation asymmetry measures can be found in the supplementary material.

\begin{figure*}
    \centering
    \includegraphics[width=\linewidth]{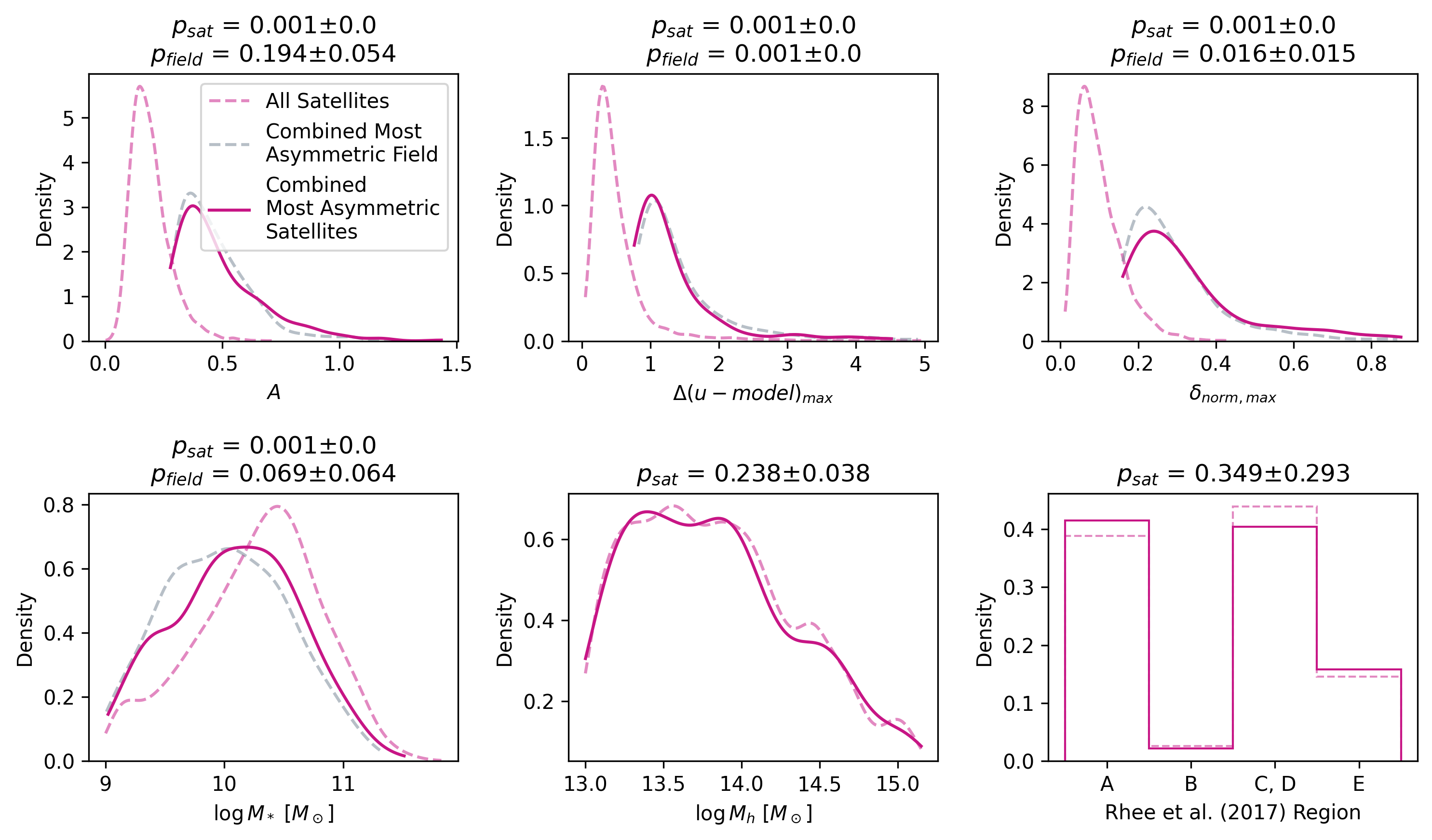}
    \caption{KDE distributions of asymmetry, maximized deviation from the model, maximized normalized difference, stellar mass, and halo mass, and a histogram of \protect\cite{rhee_2017}. Distributions are shown for the combined maximized asymmetry subsample (solid magenta lines), the total satellite sample (dashed magenta lines), and the combined maximized asymmetry field subsample (dashed grey lines). p-values and their $1\sigma$ errors comparing the two satellite samples ($p_{sat}$) and the highly asymmetric satellite and field samples ($p_{field}$) are given above each panel p-values are computed using the AD test for continuous distributions and the chi-square test for discrete distributions.}
    \label{figure_7}
\end{figure*}

Finally, we create a combined asymmetry measure by selecting galaxies that are within the top $20\%$ most asymmetric in three measures. For this, we choose to use $A$, $\Delta(u-model)_{max}$, and $\delta_{norm,max}$, as these were the most reliable measures when studying the Coma Cluster jellyfish sample in Section \ref{section_comacluster} (discussed in Section \ref{section_discussion_coma}) and visually were most distinct in Figure \ref{figure_5}. We obtain a sample of 366 galaxies, representing $\sim7\%$ of the total satellite sample. In Figure \ref{figure_7}, we compare the properties of the combined most asymmetric subsample, and the full satellite sample (with the subsample removed). Unlike in Figure \ref{figure_6}, the distributions in the star formation asymmetry measures overlap, which is expected as we are not necessarily selecting the galaxies with the highest asymmetry in all measures. Instead, we are looking broadly at which galaxies would be selected using several measures. Again, we control for stellar mass effects by applying the same cuts to the field sample to create a most asymmetric field subsample of 491 galaxies.

We compute p-values with the same method as for Figure \ref{figure_6}, given above each panel in Figure \ref{figure_7}. The most asymmetric galaxies do not favour any particular environment, with similar distributions in host halo mass and PPS region when compared to the overall satellite population. We also see that the most asymmetric satellite and field galaxies are both skewed to slightly lower stellar masses.

\section{Discussion}
\label{section_5}

\subsection{Significance of the Coma Cluster sample}
\label{section_discussion_coma}

We initially tested our methodology by applying a range of asymmetry measures to a sample of jellyfish identified in the Coma Cluster. It is worth noting, however, that \cite{roberts_2021a} showed that the Coma Cluster has a high fraction of jellyfish as compared to other low redshift clusters, meaning that the Coma Cluster may be a special environment in which to test our methods. The higher fraction may simply be due to Coma's relatively high halo mass, introducing not only more opportunities for galaxies to become jellyfish, but for higher stellar mass galaxies to become jellyfish. Moreover, merger activity in the Coma Cluster, as indicated by the presence of two central cD galaxies \citep[NGC 4874 and NGC 4889;][]{fitchett_1987}, as well as recent measurements in the radio continuum \citep{bonafede_2010,brown_2011} and the x-ray \citep{lyskova_2019,churazov_2023}. Mergers can increase the strength of ram pressure through ICM motions and high velocity dispersions.

In Figure \ref{figure_2}, both visually and with the AD test, we see that our star formation asymmetry measures are detecting most of the Coma Cluster jellyfish, as the jellyfish distributions are skewed to high asymmetries, and the AD test p-values are typically $<0.05$. However, we must consider that the sample sizes are still small, (32 jellyfish from \cite{roberts_2020} and 25 jellyfish from \cite{roberts_2021a}). Although we are number and mass-matching the samples to the control Coma Cluster sample, there are not enough normal star-forming galaxies within Coma with UNIONS $u$-band imaging for us to mass-match 100 times as we have done in the rest of our analysis. Thus, in this case, we do not have a distribution of p-values.

We can also begin to distinguish which of our methods are most reliable. First, our direction-dependent methods $\delta_{norm,cent}$ and $\Delta(u-model)_{cent}$ are not able to distinguish the jellyfish galaxies from the rest of the Coma Cluster. This may be due to our assumption of direction of motion given that the corresponding maximized measure is able to cut the galaxies in half in such a way that the jellyfish are distinct from the rest of the population. Shape asymmetry is unable to distinguish the jellyfish galaxies in the \cite{roberts_2021a} sample, which could be due to the slightly smaller sample size as compared to the \cite{roberts_2020} jellyfish sample. Using Coma jellyfish as our test population, $A$, $\delta_{norm,max}$, and $\Delta(u-model)_{max}$ are the most reliable measures of star formation asymmetry, keeping $A_s$ and $\Delta(u-r)_{max}$ as secondary measures.

\subsection{Comparison to the field sample}
\label{subsection_comparion_to_field}

In Section \ref{section_asymmetry_distributions}, we presented the distributions of asymmetry measures for our sample of group and cluster galaxies as compared to the control field sample. In most cases, the samples have very similar distributions in asymmetries as confirmed using the AD test. Only some asymmetry measures have subtle differences in their distributions when comparing the full samples ($A$, $A_s$, $\delta_{norm,max}$, $\Delta(u-model)_{max}$), but by visual examination, these differences appear to be caused by higher values in the field sample (Figure \ref{figure_3}). This could be because galaxies in dense environments are more truncated and thus show less asymmetry than the field due to the short timescale of RPS.

We further dissected the sample to examine if there are differences in distributions when dividing by galaxy and environmental properties, as ram pressure preferentially affects low stellar mass galaxies during their first infall towards high mass clusters (Figure \ref{figure_4}). However, we still do not find any significant differences. Using the AD test, we find that most satellite distributions are not distinct from those of the field, and where distinct, the differences are subtle and appear to be caused by slightly higher asymmetries in the field sample. We also note that the phase space region where we would expect ram pressure to be strongest (region B in \cite{rhee_2017}) is not very populated (only $\sim2.7\%$ of the total sample). It may be difficult to find galaxies with enhanced star formation simply because there are not enough galaxies at this point in their evolution. This small fraction is not due to any choices made when creating our sample: this region of phase space is simply not very populated. We also compared our asymmetry measures using the PPS zones defined in \cite{pasquali_2019} and still found no dependence on time since infall.

With all of this in hand, we can say that our tests indicate no significant statistical difference between the satellite sample and the field sample overall and that our results do not indicate any statistical asymmetric enhancement of star formation in infalling galaxies.

We find our results are broadly consistent with previous studies. In a study of $g-r$ colour on the leading and trailing sides of infalling group galaxies, \cite{rodriguez_2020} show that the difference in $g-r$ colour between halves, assuming the galaxies are falling directly towards the group centre in projection, is consistent with 0 on average (see Figure 6 in \cite{rodriguez_2020}). For galaxies beyond $0.75r_{180}$ that have their semi-major axes perpendicular to the projected group centre direction, they find that the leading side is slightly bluer than the trailing side in $g-r$. We do not find this small effect using our $u-r$ colours.

In simulation work, \cite{troncosoiribarren_2020} find that the cut which maximizes star formation asymmetry is well correlated with the true direction of motion of a galaxy, and is thus the best observational method to study ram pressure effects. They find no difference in SFR between halves when using a cut perpendicular to the group/cluster centre direction, consistent with our findings. Using the cut which maximizes star formation asymmetry in 3D, they find that satellite galaxies have slightly higher SFR differences as compared to the field, but it is unknown how significant this difference is given that the spread in the field values is not shown. Using leading and trailing halves defined by the 3D velocity vector, \cite{troncosoiribarren_2020} find an SFR difference of $\sim10\%$ between leading and trailing halves. Similarly, our median value of $\delta_{norm,max}$ is $\sim9\%$, however, our median field value is also $\sim9\%$, while their median field value using random cuts is $\sim0$. This aligns with the fact that \cite{troncosoiribarren_2020} find that SFR excess in satellites is only slightly higher than in the field when using the cut which maximizes star formation asymmetry.

Studies of jellyfish galaxies \citep[e.g.][]{roberts_2022,roberts_2023} have found star formation anisotropies when comparing leading and trailing halves. However, these studies are of galaxies with observable tails of gas, providing the direction of motion. This is likely not the primary reason why we are unable to detect a significant asymmetric star formation enhancement, as our `$max$' methods would be able to provide a proxy for the true direction of motion.

\subsection{What causes high star formation asymmetries?}
\label{section_causes}

When restricting the sample to galaxies which have low stellar mass, high halo mass, and small time since infall (Figure \ref{figure_5}), we find that the subsample of galaxies likely experiencing strong ram pressure and the full satellite sample are statistically distinct using the AD test in most cases. However, the subsample of galaxies likely experiencing strong ram pressure is only slightly skewed towards higher star formation asymmetries, and there is no clear cut that could distinguish galaxies experiencing enhanced star formation from those that are not. It is also important to note that the \cite{rhee_2017} regions are contaminated, and while it would be ideal to restrict our sample to the galaxies in the region where RPS is truly the strongest (region B), this region contains very few galaxies, and our RPS-likely sample would reduce to only 24 galaxies. Thus, we choose to include region A, which contains the highest fraction of first infallers, despite the fact that it is highly contaminated with interlopers ($>50\%$). In addition, the strong ram pressure subsample is indistinguishable from the low stellar mass field sample in each measure, suggesting that the slightly higher asymmetries are likely driven by stellar mass effects. This could indicate that dense environments act to reduce asymmetries over longer timescales, with any enhanced star formation potentially being a brief stage in this process.

We finally investigate the properties of the most asymmetric galaxies in the satellite sample (Figure \ref{figure_6}). We see that the stellar mass distributions are skewed to low-mass galaxies by-eye as expected but these differences are only statistically significant in the galaxies with high $A$. The stellar mass distributions of the most asymmetric field galaxies are also skewed toward low stellar mass, again pointing towards a stellar mass effect.

To create a more powerful measure of star formation asymmetry, we combined three of our measures, $A$, $\Delta(u-model)_{max}$, and $\delta_{norm,max}$, as discussed in Section \ref{section_discussion_coma} (Figure \ref{figure_7}). The most asymmetric galaxies in groups and clusters are preferentially lower mass galaxies but have no environmental dependence.

\section{Summary and Conclusions}
\label{section_6}

In this work, we study the effect of ram pressure on star formation in galaxies falling into groups and clusters. We use UNIONS $u$-band imaging as a tracer of SFR and measure asymmetry in star formation using several direction-dependent and independent measures. 

We find no dependence of star formation asymmetry on environment when comparing galaxies living in groups and clusters to galaxies living in isolation. We further divide the satellite sample into bins of stellar mass, halo mass, and time since infall, and still find no dependence of star formation asymmetry on galaxy properties. When restricting our sample to galaxies likely experiencing strong ram pressure, we see slightly higher asymmetries, but these differences are minimal. Similarly, when limiting our sample to the galaxies with the highest asymmetries, we see that galaxies preferentially have some of the expected properties, but these differences are insignificant and also observed in the field. This could also indicate that the group and cluster environments act to smooth asymmetries over time, and any enhancement is potentially a brief stage in this process.

Thus, we conclude that any statistical ram pressure-induced star formation enhancement for infalling galaxies is small. This means that this phenomenon is either uncommon and \textit{rarely occurs}, or that this effect is short-lived, and is simply \textit{rarely observed}. It is outside the scope of this work to constrain the timescale of this effect. Observationally, this would be a difficult task due to the scarcity of jellyfish galaxies in the local universe, and the challenges of assigning time since infall to group and cluster galaxies.

Future work could include using cosmological simulations (e.g., IllustrisTNG; \cite{nelson_2019}) to study and constrain the timescale of ram pressure-induced star formation enhancement on infalling galaxies. As discussed throughout, previous works have studied the effect of ram pressure on star formation using simulations \citep[e.g.][]{troncosoiribarren_2020,oman_2021,goller_2023,zhu_2024}, but none have focused on constraining the timescale of the star formation enhancement. \cite{goller_2023} studied the SFR in jellyfish galaxies in IllustrisTNG, and found that while $74\%$ of jellyfish experienced a starburst at some point in their history, this did not translate to a population-wide enhancement. This motivates a similar analysis on normal satellites, as well as a resolved analysis. Our study could also be improved by using higher-quality SFR maps. We tested our sample using H$\alpha$ maps from the MaNGA survey \citep{bundy_2015}, which is a better tracer of SFR than the UNIONS $u$-band imaging. However, we were unable to draw robust conclusions due to the small sample size and the limited radial extent of the data.

\section{Acknowledgements}

We thank the anonymous referee for their useful and insightful comments. LMF thanks the Ontario Graduate Scholarship program for funding, LMF and LCP thank the Natural Science and Engineering Research Council for funding, and IDR acknowledges support from the Banting Fellowship Program.

We are honoured and grateful for the opportunity to observe the Universe from Maunakea and Haleakala, which both have cultural, historical and natural significance in Hawaii. This work is based on data obtained as part of the Canada-France Imaging Survey, a CFHT large program of the National Research Council of Canada and the French Centre National de la Recherche Scientifique. This research used the facilities of the Canadian Astronomy Data Centre operated by the National Research Council of Canada with the support of the Canadian Space Agency.

\vspace{5mm}
\facilities{CFHT}

\software{\texttt{astropy} \citep{astropy_2013,astropy_2018,astropy_2022}, \texttt{imfit} \citep{erwin_2015}, \texttt{matplotlib} \citep{hunter_2007}, \texttt{numpy} \citep{harris_2020}, \texttt{pandas} \citep{mckinney_2010, reback_2020}, \texttt{photutils} \citep{bradley_2024}, \texttt{scipy} \citep{virtanen_2020}, \texttt{seaborn} \citep{waskom_2021}, and \texttt{topcat} \citep{taylor_2005}.}

\appendix
\section{Data Table}
\label{appendix}
We include a machine-readable data table of the galaxy ($M_*$, SFR, position angle $\theta$, ellipticity $e$, effective radius $r_e$, SNR, and inclination angle $i$) and environmental ($M_h$, PPS region) properties used in this analysis, as well as our measured asymmetry values. Table \ref{table1} shows an example of 5 rows of satellite galaxies with both $u$ and $r$-band imaging.

\begin{splitdeluxetable}{cccccccccBccccccccccBccccc}
\label{table1}
\tablecaption{Galaxy intrinsic and environmental properties and their star formation asymmetries}
\tablehead{
    \colhead{RA} & \colhead{Dec.} & \colhead{$z$} & \colhead{Type} & \colhead{$\log M_* [M_\odot]$\tablenotemark{a}} & \colhead{$\log \text{SFR} [M_\odot \text{yr}^{-1}]$\tablenotemark{a}} & \colhead{Group ID\tablenotemark{b}} & \colhead{$\log M_h [M_\odot]$\tablenotemark{b}} & \colhead{Phase Space Region\tablenotemark{c}} & \colhead{$\theta_u [\text{rad}]$\tablenotemark{d}} & \colhead{$e_u$} & \colhead{$r_{e,u} [\text{"}]$} & \colhead{$\text{SNR}_u$} & \colhead{$i_u [\degree]$} & \colhead{$\theta_r [\text{rad}]$\tablenotemark{d}} & \colhead{$e_r$} & \colhead{$r_{e,r} [\text{"}]$} & \colhead{$\text{SNR}_r$} & \colhead{$i_r [\degree]$} & \colhead{$\delta_{norm,max}$} & \colhead{$\Delta(u-model)_{max}$} & \colhead{$\Delta(u-r)_{max}$} & \colhead{$A$} & \colhead{$A_s$}
} 
\startdata 
    119.416 & 40.946 & 0.074 & Satellite & 11.0 & -0.093 & 4718 & 13.55 & C & 1.57 & 0.04 & 2.183 & 315.0 & 16.23 & 1.57 & 0.064 & 3.413 & 3601.0 & 20.6 & 0.034 & 0.229 & -0.078 & 0.085 & 0.385 \\
    117.152 & 39.002 & 0.085 & Satellite & 10.39 & -0.039 & 8551 & 13.33 & C & 3.05 & 0.112 & 1.229 & 202.0 & 27.34 & 1.57 & 0.0 & 1.412 & 1873.0 & 0.0 & 0.127 & 0.824 & -0.153 & 0.303 & 0.38 \\
    117.512 & 39.507 & 0.097 & Satellite & 10.43 & 0.05 & 3073 & 13.69 & D & 1.57 & 0.339 & 2.495 & 183.0 & 48.66 & 1.57 & 0.504 & 2.272 & 1862.0 & 60.23 & 0.091 & 0.137 & -0.068 & 0.122 & 0.415 \\
    118.791 & 41.139 & 0.075 & Satellite & 10.58 & 0.356 & 728 & 14.01 & D & 0.63 & 0.047 & 3.781 & 464.0 & 17.58 & 1.57 & 0.0 & 3.784 & 3222.0 & 0.0 & 0.11 & 0.379 & -0.12 & 0.2 & 0.529 \\
    119.044 & 41.153 & 0.073 & Satellite & 10.78 & 0.024 & 728 & 14.01 & C & 1.19 & 0.109 & 2.192 & 372.0 & 26.94 & 1.34 & 0.126 & 2.295 & 3173.0 & 29.05 & 0.059 & 0.845 & -0.037 & 0.208 & 0.291 \\
\enddata
\tablenotetext{a}{From the GSWLC-M2 \citep{salim_2016,salim_2018}}
\tablenotetext{b}{From the \cite{yang_2007} catalogue}
\tablenotemark{c}{\cite{rhee_2017} PPS region}
\tablenotetext{d}{Counterclockwise from the positive x-axis}
\end{splitdeluxetable}

\bibliography{biblio}{}
\bibliographystyle{aasjournal}

\end{document}